\documentstyle[prb,aps,epsfig]{revtex}
\begin{document}
\draft
\newcommand{\be}{\begin{equation}}
\newcommand{\ee}{\end{equation}}
\newcommand{\ba}{\begin{eqnarray}}
\newcommand{\ea}{\end{eqnarray}}
\newcommand{\vk}{{\bf k}}
\newcommand{\vq}{{\bf q}}
\newcommand{\vp}{{\bf p}}
\newcommand{\vx}{{\bf x}}

\def\inseps#1#2{\def\epsfsize##1##2{#2##1} \centerline{\epsfbox{#1}}}

\def\top#1{\vskip #1\begin{picture}(290,80)(80,500)\thinlines \put(65,500){\line( 1, 0){255}}\put(320,500){\line( 0, 1){
5}}\end{picture}}
\def\bottom#1{\vskip #1\begin{picture}(290,80)(80,500)\thinlines \put(330,500){\line( 1, 0){255}}\put(330,500){\line( 0, -1){
5}}\end{picture}}


\title{The Influence of  Quantum Critical Fluctuations of Circulating Current Order Parameters
on the  Normal State Properties of Cuprates}
\author{Hyun C. Lee and Han-Yong Choi}
\address{BK21 Physics Research Division and Institute of Basic Science,
Department of Physics,\\
Sung Kyun Kwan University, Suwon 440-746 Korea}
\maketitle
\draft
\begin{abstract}
We study a model of the quantum critical point of cuprates
associated with the "circulating current" order parameter
proposed by Varma. An effective action of the order parameter in
the quantum disordered phase is derived using functional integral
method, and the physical properties of the normal state are
studied based on the action. The results derived within the ladder
approximation indicate that the system is like Fermi liquid near
the quantum critical point and in disordered regime up to minor
corrections.
This implies that the suggested marginal Fermi liquid behavior
induced by the circulating current fluctuations will come in from
beyond the ladder diagrams.
\end{abstract}
\thispagestyle{empty}
\section{Introduction}
In spite of the intensive studies of the last 15 years the
microscopic origin of the high temperature superconductors (HTSC)
still remains elusive. However, the physical properties in each
regime of phase diagram are now well established. \cite{review}
The normal state properties of the cuprate supercondcutors are
highly anomalous. They deviate substantially from the Fermi
liquid behavior and are believed to be of a non-Fermi liquid.
Especially, the anomalous normal state properties near optimal
doping concentration can be characterized by the phenomenological
marginal Fermi liquid (MFL) theory. \cite{mfl} The MFL theory
rests on a single hypothesis that there exist excitations whose
spectral function is given by \be \label{marginal1} {\rm Im}
P(\omega,\vq) \sim \cases{ -N(E_F) \frac{\omega}{T}, \quad
|\omega| \ll T \cr -N(E_F) {\rm sgn} \omega, \quad |\omega| \gg
T, } \ee where $N(E_F)$ is the unrenormalized density of states
at the Fermi energy, and ${\rm sgn \omega} = \omega/|\omega|$. The
important features of the spectral function Eq.\ (\ref{marginal1})
are its independence of the momentum $\vq$ and its scaling form as
a function of $\omega/T$. The $\omega/T$ scaling is the hallmark
of quantum critical point (QCP), thus the scaling form of the
spectral function very strongly suggests a proximate QCP. If QCP
indeed underlies the MFL behaviour, the momentum independence of
Eq.\ (\ref{marginal1}) implies infinitely large dynamical
exponent $z_d$ of QCP.

Most of the normal state properties associated with two-particle
correlations near optimal doping have been found consistent with
the MFL theory. \cite{mfl} Recently, the validity of MFL was
confirmed also for single-particle properties. The recent
angle-resolved photoemission experiment by Valla {\it et al.}
\cite{valla} found that the single-particle self-energy is given
by the MFL form. They measured the imaginary part of the electron
self-energy of the optimal Bismuth HTSC along a nodal direction,
and obtained ${\rm Im} \Sigma(\vk, \epsilon) \sim \max(
|\epsilon|, T)$. This observation renewed the interest in the MFL
theory, and especially, in the microscopic basis of the MFL and
conjectured proximate QCP.

There are several models rooted on the idea of QCP which attempt
to explain various aspects of HTSC\cite{varma1,varma2,theory}. In
general, it is necessary to introduce a certain local order
parameter to define a QCP. The present experimental data put
strong constraint on the possible form of the order parameter. In
particular,  there is no evidence of broken translational
symmetry, so the order parameter of QCP should be chosen to
respect  the translational symmetry. The theory by C. M. Varma
\cite{varma1,varma2} is based on the order parameter which does
respect the translation symmetry but breaks  the time reversal
symmetry and the four-fold rotation symmetry. The construction of
such order parameter necessitates the full set of the copper $d$
and the oxygen $p_{x,y}$ orbitals. The explicit form of the
(complex) order parameter is \be \label{order} \Psi(T,
x)=\frac{V}{2} \sum_{\vp,\sigma} \langle \sin\frac{p_x a}{2}
d^\dag_{\vp \sigma} p_{x \vp \sigma}- \sin\frac{p_y a}{2}
d^\dag_{\vp \sigma} p_{y \vp \sigma} \rangle, \ee where $V$ is
coupling constant, and $a, x$ is lattice constant of unit
cell\cite{comment1} and doping concentration, respectively.
$\sigma$ is spin quantum number. At zero temperature, $\Psi(T=0,
x < x_c)\neq 0$. The critical doping $x_c \sim 0.2$
\cite{varma2,comment2} defines the QCP.

In QCP theory of HTSC by Varma, the anomalous normal state
corresponds to the finite temperature region of the disordered
regime $T > T_c > 0, x \sim x_c$. In this regime the critical
fluctuations of order parameter determine the physical properties.
Varma made an extensive study on the general class of models
which can give rise to the order parameter (\ref{order}), and
gave an heuristic (and essentially non-perturbative) derivation
of the susceptibility $\chi(\omega,\vq)$ which controls the
critical fluctuation of order parameter in the normal
state\cite{varma1}. In this paper, we have chosen a simplified
model considered by Varma \cite{varma2} and derived an effective
action (See Eq.(\ref{action1}).) of the order parameter
systematically in the functional integral formalism. Such
effective action can be also regarded as time-dependent
Landau-Ginzburg type free energy. The lowest order term of the
effective action (See Eq.(\ref{quadratic}).) amounts to the
summation of the ladder-type diagrams. The physical properties
deduced from the lowest order term of the effective action are
{\it not} compatible with the experimental data of the normal
states. For an example, the electron self-energy derived from the
lowest order effective action is Fermi liquid-like apart from
logarithmic correction. If we accept the heuristic derivations by
Varma, the inadequacy of the lowest order approximation implies
the importance of higher order terms beyond the ladder diagrams.

This paper is organized as follows: In Sec.II, we introduce our
model and derive the functional integral representation of the
effective action. In Sec.III, the concrete form of the effective
action is obtained up to the 4th order. In Sec.IV, we analyze the
effective action following Moriya's self-consistent
renormalization scheme, and calculate some physical quantities.
In Sec.V, we conclude this paper with the summary and some
concluding remarks.

\section{Formulation}
The basic model is the three-band Hubbard model\cite{varma1,varma2,varma3}.
\ba
H&=&H_0+H_1+H_2, \nonumber \\
H_0&=& \sum_{\vk,\sigma}\,\Big( \epsilon_d\,n_{d \vk \sigma}+
2 t_{pd} d^\dag_{\vk \sigma}\big[ s_x(\vk) p_{x \vk \sigma}+ s_y(\vk)
p_{y \vk \sigma} \big]
- 4 t_{pp} s_x(\vk) s_y(\vk)  p^\dag_{x \vk \sigma}
p_{y \vk \sigma}+{\rm H.C.}\Big), \nonumber \\
H_1&=&\sum_{i \sigma}\,U_d n_{d i \sigma} n_{d i -\sigma}
+
U_p ( n_{p x i \sigma} n_{p x i -\sigma} + n_{p y i \sigma} n_{p y i -\sigma}),
\nonumber \\
H_2&=& 2 V \sum_{\vk,\vk^\prime,\vq,\sigma,\sigma^\prime}\,
c_x(\vq) \, d^\dag_{\vk+\vq \sigma} d_{\vk \sigma}\,
p^\dag_{x \vk^\prime-\vq \sigma^\prime} p_{x \vk^\prime \sigma^\prime}
+
c_y(\vq) \, d^\dag_{\vk+\vq \sigma} d_{\vk \sigma}\,
p^\dag_{y \vk^\prime-\vq \sigma^\prime} p_{y \vk^\prime \sigma^\prime}.
\ea
$i$ is a lattice site index, and $c_{x,y}(\vk)=\cos \frac{k_{x,y} a}{2},
s_{x,y}(\vk)=\sin \frac{k_{x,y} a}{2}$. $H_1$ is the on-site repulsion term which suppresses
the doubly occupied states. $H_2$ is the nearest neighbor repulsion ($V>0$) term between copper $d$ and
oxygen $p_x,p_y$ orbitals written in momentum space.
The order parameter of circulating current (CC)  phase comes from $H_2$. To simplify our problem,
we implement the effect of $H_1$
by a substitution $ t_{pd} \to x t_{pd},\;
t_{pp} \to x t_{pp} $\cite{varma2}. Certainly, this is very crude approximation and the effect
of $H_1$ shoud be more carefully included in  due course.

It is easy to show,  by re-indexing momenta,
that $H_2$ can be written as
\ba
\label{hubbard}
H_2&=&-V \sum_{i=1,2,3,4 \sigma \sigma^\prime}
\hat{O}^\dag_{\sigma \sigma^\prime i}  \hat{O}_{\sigma^\prime \sigma i} \nonumber \\
\hat{O}_{1 \sigma^\prime \sigma}(\vq)&=&
\sum_{\vk^\prime}\,\Big( s_{\vk^\prime x} p^\dag_{x \vk^\prime \sigma^\prime}
- s_{\vk^\prime y} p^\dag_{y \vk^\prime  \sigma^\prime}\Big)\,d_{\vk^\prime+\vq \sigma}, \nonumber \\
\hat{O}_{2  \sigma^\prime \sigma}(\vq)&=&
\sum_{\vk^\prime}\,\Big( s_{\vk^\prime x} p^\dag_{x \vk^\prime \sigma^\prime}
+ s_{\vk^\prime y} p^\dag_{y \vk^\prime  \sigma^\prime}\Big)\,d_{\vk^\prime+\vq \sigma}, \nonumber \\
\hat{O}_{3  \sigma^\prime \sigma}(\vq)&=&
\sum_{\vk^\prime}\,\Big( c_{\vk^\prime x} p^\dag_{x \vk^\prime \sigma^\prime}
- c_{\vk^\prime y} p^\dag_{y \vk^\prime  \sigma^\prime}\Big)\,d_{\vk^\prime+\vq \sigma}, \nonumber \\
\hat{O}_{4  \sigma^\prime \sigma}(\vq)&=&
\sum_{\vk^\prime}\,\Big( c_{\vk^\prime x} p^\dag_{x \vk^\prime \sigma^\prime}
+ c_{\vk^\prime y} p^\dag_{y \vk^\prime  \sigma^\prime}\Big)\,d_{\vk^\prime+\vq \sigma}.
\ea
The expectation value of the operator $\hat{O}_{1 \sigma \sigma}\equiv \hat{O}_{1 \sigma} $
is the order parameter of
CC phase Eq.(\ref{order}) up to constant factor. Because  the critical fluctuation of
$\hat{O}_{1 \sigma}$ is most dominant near QCP, the remainder
$H_2^\prime=-V \sum_{i=2,3,4,\sigma^\prime \sigma}
\hat{O}^\dag_{\sigma^\prime i}  \hat{O}_{\sigma i}-V \sum_{\sigma^\prime \neq \sigma}
\hat{O}^\dag_{\sigma^\prime \sigma 1}  \hat{O}_{\sigma \sigma^\prime 1}$
can be treated as perturbations near QCP.
We will not take the effect of $H_2^\prime$ into account explicitly in this work.
Note that the operator $\hat{O}_{1 \sigma}$ is {\it odd} under the $C_4$ symmetry of square
lattice.
The reduced Hamiltonian which incoporates the physics associated with
CC phase is
\ba
\tilde{H}&=&
\sum_{\vk,\sigma}\,\Big( \epsilon_d\,n_{d \vk \sigma}+
2 t d^\dag_{\vk \sigma}\big[ s_x(\vk) p_{x \vk \sigma}+ s_y(\vk)
p_{y \vk \sigma} \big]
 - 4 t^\prime s_x(\vk) s_y(\vk)  p^\dag_{x \vk \sigma}
p_{y \vk \sigma}+{\rm H.C.}\Big) \nonumber \\
&-&V \sum_{\sigma, \vq} \hat{O}_\sigma^\dag(\vq)\,
\hat{O}_\sigma(\vq),
\ea
where $t\equiv x t_{pd}, t^\prime \equiv x t_{pp}$.
We could have adopted a basis where kinetic term of the reduced Hamiltonian is diagonalized.
The advantage of our choice of  basis is
that the coupling between  order parameter and  fermions is
simple and that it admits the formal manipulations of  functional integral method.
Using  Hubbard-Stranovich transformation,
the partition function  can be
written in the functional integral form.
\ba
\label{partition}
Z&=& \int D[\Psi_{l \sigma},\bar{\Psi}_{l \sigma^\prime}]\,D[N_\sigma]\,
\exp[-S(\Psi,\bar{\Psi},N)],\nonumber \\
S&=&\sum_{lm,\sigma \sigma^\prime}\, \sum_{i\epsilon, \epsilon^\prime,
\vk, \vk^\prime}\,
\bar{\Psi}_{l \sigma}(i\epsilon^\prime,\vk^\prime)
\,\Big[ K_0^{-1}-V M \Big]\,\Psi_{m \sigma^\prime}(\epsilon,\vk)
+
V \sum_\sigma \sum_{\vq,\omega}  N_\sigma^\dag(i\omega,\vq) N_\sigma(i\omega,\vq),
\ea
where the six component spinor is defined
$\Psi_{l \sigma}(i\epsilon,\vk)
=\Big( d_{\vk \sigma}, p_{x \vk \sigma}, p_{y \vk \sigma} \Big)$.
$l,m$ are the orbital indices.
The explicit form of  kernel matrix is ($\mu$ is the chemical potential.)
\be
\label{kernel}
K_0^{-1}(i\epsilon,i\epsilon^\prime;\vk,\vk^\prime;lm;\sigma \sigma^\prime)=
\delta_{\epsilon \epsilon^\prime}\,\delta_{\sigma \sigma^\prime} \delta_{\vk \vk^\prime}
\pmatrix{\epsilon_d  -i\epsilon-\mu & 2 t s_x(\vk) & 2 t s_y(\vk)  \cr
          2 t s_x(\vk) & -i \epsilon-\mu & -4 t^\prime s_x(\vk) s_y(\vk) \cr
      2 t s_y(\vk) &  -4 t^\prime s_x(\vk) s_y(\vk) & -i \epsilon-\mu}.
\ee
The explicit form of the matrix $M$ is ($k=(i\epsilon,\vk), k^\prime=(i \epsilon^\prime,\vk^\prime)$.)
\be
\label{theM}
M(k^\prime, k; \sigma^\prime \sigma;lm)=\delta_{\sigma \sigma^\prime}
\pmatrix{ 0 & N_{k^\prime-k} s_x(\vk) & -N_{k^\prime-k} s_y(\vk) \cr
          N^\dag_{k-k^\prime}s_x(\vk^\prime)&   0  & 0  \cr
      -N^\dag_{k-k^\prime} s_y(\vk^\prime) & 0 & 0 }.
\ee
The effective action of order parameter $N_\sigma$ can be obtained by
integrating out the electrons $\Psi_{\sigma l}$.
($q=(i\omega,\vq)$.)
\be
\label{action0}
S_{{\rm eff}}=V \sum_{\sigma q} N^\dag_\sigma(q)  N_\sigma(q)-2 \,\ln
{\rm Det}(K_0^{-1}-V M),
\ee
where the factor of 2 of the second term comes from the spin degeneracy. From now on the
spin index of $N$ will be suppressed. By expanding the determinant of (\ref{action0})
the (time-dependent) Landau-Ginzburg type action can be obtained.
In principle, the action (\ref{action0}) contains
all the elements for the description of the critical properties of our model.
The mean-field phase diagram  can be determined by minimizing
(\ref{action0}) with respect to $N, N^\dag$.
Using  relation $\delta \ln \det A=\delta {\rm Tr} \ln A={\rm Tr} A^{-1} \delta A$, we get
\be
\label{mfe}
N^\dag(q)=-2\sum_{k,k^\prime,a,b}
\left(\frac{1}{K_0^{-1}-VM}\right)_{k,k^\prime; lm}\,
\frac{\delta M_{k^\prime, k; ml}}{\delta N(q)}.
\ee
Eq.(\ref{mfe}) is the mean-field equation written in matrix form.
According to the mean-field solution by Varma\cite{varma2},  Eq.(\ref{mfe}) has a non-trivial
solution for $x < x_c \sim 0.2$ and $ T< T_c(x)$ with $T_c(x_c)=0$.
The non-trivial mean-field solution is
of the form $N= i R$, where $R$ is real.  It should be noted that only the solution of this type
preserves $C_4 \times T$, where $T$ is the time reversal.
 Thus, we expect that the susceptibility of the
imaginary component of $N$
$\langle {\rm Im} N (q) {\rm Im} N(-q) \rangle$
would diverge as $q \to 0, x \to x_c + 0^+$ at zero temperature,
reflecting the large critical
fluctuation near QCP. The susceptibility determines the critical properties near QCP.
In the quantum disordered phase, the mean-field equation (\ref{mfe}) has a trivial solution $N=0$, and
we may expand the action (\ref{action0}) as a power series of $N$.

\section{Effective Action}
One can show that the terms with odd powers of $N$ vanish in the expansion of
the determinant of (\ref{action0}) by using the transformation property
of $N$ under  $C_4$ rotation.
Expanding the determinant up to the 4-th order we get
\ba
\label{action1}
S_{{\rm eff}}&=&2 V\sum_{ q} N^\dag(q) N(q) +
\sum_q \,\Big[ \Pi_A(q)\,N(q) N^\dag(q) + \Pi_B(q)\,(N(q)N(-q)+N^\dag(q) N^\dag(-q))\Big]
\nonumber \\
&+&\sum_{q_i,i=1,2,3,4}\,\Big[ \Pi_C(q_i) N(q_1) N(q_2) N^\dag(q_3) N^\dag(q_4)+
\Pi_D(q_i)\,N(q_1) N^\dag(q_2)\,(N(q_3) N(q_4)+N^\dag(q_3) N^\dag(q_4)) \nonumber \\
&+&
\Pi_E(q_i)\, (N(q_1)N(q_2)N(q_3)N(q_4)
+N^\dag(q_1)N^\dag(q_2)N^\dag(q_3)N^\dag(q_4)) \Big]+O(N^6,N^8,\ldots).
\ea
The appropriate momentum-energy conservation rules are to be understood for the 4th order terms.
The explicit expressions of the polarization functions
$\Pi_i$ are relegated to the appendix. Since the $d, p_x, p_y$ electrons are
the linear combinations of the anti-bonding $a$, the bonding $b$, and the non-bonding $c$
electron operators, the polarization functions $\Pi_i$'s contain both intra-band ($a a$) and
inter-band ($a b$ or $a c$) contributions.

For analyses later, it is convenient to decompose the order parameter  $N$ into the real
and imaginary components.
In real space ($x=(\tau,{\bf x})$.), the decomposition is
\be
N(x)=N_1(x)+i N_2(x), N^\dag(x)=N_1(x)-i N_2(x).
\ee
In momentum-frequency space,
\be
N(q)=N_1(q)+ i N_2(q),
N^\dag(q)=N_1(-q)-i N_2(-q).
\ee
In terms of real components $N_1, N_2$ the quadratic part of the effective action (\ref{action1})
can be written as
\be
\label{oldquad}
S_{{\rm eff}}^{(2)}=\sum_q \Big[ \tilde{\pi}_1(q) N_1(q) N_1(-q) +\tilde{\pi}_2(q) N_2(q) N_2(-q) \nonumber \\
+i \tilde{\pi}_3(q) \big(N_1(q) N_2(-q)-N_1(-q) N_2(q) \big) \Big],
\ee
where
\ba
\tilde{\pi}_1(q)&=&2 V+\Pi_A(q)+2 \Pi_B(q), \nonumber \\
\tilde{\pi}_2(q)&=&2 V+\Pi_A(q)-2 \Pi_B(q), \nonumber \\
\tilde{\pi}_3(q)&=&2 V+\Pi_A(q).
\ea
Using the results Eq.(\ref{polarization}) for $\Pi_A(q)$ and $\Pi_B(q)$,
 $\tilde{\pi}_i(q)$'s ($i=1,2,3$) can be parametrized as
\be
\label{pipol}
\tilde{\pi}_i(i \omega,\vq)= \delta_i + a_i |\vq|^2 + b_i \omega^2+
\frac{ |\omega|}{\Gamma_i |\vq|}-
i \eta \omega,
\ee
where $a_1=a_3, \Gamma_1=\Gamma_3$. The polarization functions $\tilde{\pi}_i(q)$
contain both the intra-band and the inter-band contributions. In particular, the
Landau damping term which exists for $|\omega|<\Gamma_i |\vq|$ solely stems from the intra-band
contribution.
The  parametrizations Eq.(\ref{pipol}) are  valid for $|\omega| < \mu$ (See Appendix B.).
At this point, we observe that $N_1(q) N_1(-q)$ and $N_2(q) N_2(-q)$ of Eq.(\ref{oldquad})
are even in $q$, while $(N_1(q) N_2(-q)- N_1(-q) N_2(q))$ is
odd in $q$. Therefore, only the even parts of  $\tilde{\pi}_1(q)$ and $\tilde{\pi}_2(q)$
 and the odd part of $\tilde{\pi}_3(q)$ contribute to $S_{{\rm eff}}^{(2)}$.
\be
\label{quadratic}
S_{{\rm eff}}^{(2)}=\sum_q \Big[ \pi_1(q) N_1(q) N_1(-q) +\pi_2(q) N_2(q) N_2(-q)
+i \pi_3(q) \big(N_1(q) N_2(-q)-N_1(-q) N_2(q) \big) \Big],
\ee
\ba
\label{s20}
\pi_1(i \omega,\vq)&=& \delta_1 + a_1 |\vq|^2 + \frac{ |\omega|}{\Gamma_1 |\vq|}
+b_1 \omega^2,\nonumber \\
\pi_2(i \omega,\vq)&=& \delta_2 + a_2 |\vq|^2 + \frac{ |\omega|}{\Gamma_2 |\vq|}
+b_2 \omega^2, \nonumber \\
\pi_3(i \omega,\vq)&=&-i \eta \omega.
\ea
The propagators (or the susceptibilities) of order parameters can be read off from
Eq.(\ref{quadratic},\ref{s20}).
\ba
\label{susceptibility}
\langle N_1(q) N_1(-q) \rangle&=& \frac{\pi_2(q)}{\pi_1(q) \pi_2(q)-(\pi_3(q))^2}, \nonumber \\
\langle N_2(q) N_2(-q) \rangle&=& \frac{\pi_1(q)}{\pi_1(q) \pi_2(q)-(\pi_3(q))^2}, \nonumber \\
\langle N_1(q) N_2(-q) \rangle&=& \frac{-i\pi_3(q)}{\pi_1(q) \pi_2(q)-(\pi_3(q))^2}.
\ea
Let us define $D_{ij}(q)=\langle N_i(q) N_j(-q) \rangle$.
The explicit form of the $D_{22}(i\omega,\vq)$ is
\be
\label{d221}
D_{22}(i\omega,\vq)=\Big[\delta_2+a_2 |\vq|^2+b_2 \omega^2+ \frac{|\omega|}{\Gamma_2 |\vq|}+\frac{
\eta^2 \omega^2}{\delta_1+a_1 |\vq|^2+b_1 \omega^2+ \frac{|\omega|}{\Gamma_1 |\vq|}}\Big]^{-1}.
\ee
Eq.(\ref{d221}) is valid within the ladder approximation (See below.).
As discussed in the previous section, the critical fluctuation of $N_2$ is important near QCP, thus
the susceptibility $D_{22}(i\omega,\vq)$ determines the physical properties near QCP.
Obviously, $\delta_2$ is the parameter which controls the proximity  to QCP. The precise location of
QCP is determined by the condition
\be
[D_{22}^{(ren)}]^{-1}(i\omega \to 0, \vq \to 0)=0,
\ee
where $D_{22}^{(ren)}(i\omega,\vq)$ is the fully renormalized susceptibility
which includes all of the higher order corrections.
Since $\delta_1$ is expected to be large, we can approximate the last term of (\ref{d221}) as
$\frac{\eta^2}{\delta_1} \omega^2$ and absorb it into $b_2$. Then, within ladder approximation,
the imaginary part of the  retarded $D_{22}^{R}(\omega,\vq)$ becomes
\ba
\label{d22}
{\rm Im} D_{22}^{R}(\omega>0,\vq) &\approx& \Theta(\Gamma_2 |\vq|-\omega)
\frac{\frac{\omega}{\Gamma_2 |\vq|}}{(\delta_2+a_2 |\vq|^2)^2+(\frac{\omega}{\Gamma_2 |\vq|})^2}
\nonumber \\
&+& \Theta(\omega-\Gamma_2 |\vq|) \delta(\delta_2+a_2 |\vq|^2-b_2 \omega^2),
\ea
where $\Theta(x)$ is the step function ($\Theta(x)=1, x> 0; \Theta(x)=0, x< 0$.).

If we neglect the higher order terms like $N^4, N^6, \ldots$,
the quadratic action Eq.(\ref{quadratic}) amounts to the
ladder approximation. This can be verified by the direct calculation of
susceptibility $\langle \hat{O}(q) \hat{O}^\dag(q) \rangle$ (See Eq.(\ref{hubbard}).) both
by  the functional integral formalism and  the conventional perturbation theory.
Then, following  Mahan's treatment of excitons \cite{mahan}, we might expect the
excition-like poles in the susceptibility.
This pole can be identified in Eq.(\ref{d221}) by analytic continuation $i\omega \to \omega+i\delta$.
QCP in the ladder approximation is equivalent to the vanishing of the exciton-like gap.

The coefficients of the 4th order terms $\Pi_C(q_i), \Pi_D(q_i), \Pi_E(q_i)$ depend on
the momenta of  order parameters.  For small $q_i$'s, $\Pi_C(q_i), \Pi_D(q_i), \Pi_E(q_i)$
can be expanded in $q_i$.
Neglecting the dependences on momenta, the 4th order terms of the effective action
can be expressed as (in real space $x=(\tau,{\bf x})$.)
\be
\label{4thorder}
S_{{\rm eff}}^{(4)}=
\int d^3 x \Big[\frac{\lambda_1}{4!}\,N_1^4(x)+ \frac{\lambda_2}{4!} \, N_2^4(x)+
\frac{\lambda_{3}}{2! 2!}\,N_1^2(x) N_2^2(x) \Big],
\ee
where
\be
\lambda_1=4! (\Pi_C(0)+2 \Pi_D(0)+2 \Pi_E(0)),\;\;
\lambda_2=4! (\Pi_C(0)-2 \Pi_D(0)+2 \Pi_E(0)),\;\;
\lambda_{3}=2! 2! (2 \Pi_C(0)-12 \Pi_E(0) ).
\ee
$S_{{\rm eff}}^{(4)}$ introduces the lowest order corrections beyond ladder diagrams.
It also renormalizes the electron self-energy at one loop order.
The importances of still higher order terms will be discussed at Sec.V.

Near  QCP $\delta_2=0$, the naive perturbation theory breaks down because of
the infrared divergences as can be demonstrated by caculating the correction
to free energy with $S=S_{{\rm eff}}^{(2)}+S_{{\rm eff}}^{(4)}$.
These divergences can be handled by the self-consistent renormalization
scheme developed by Moriya\cite{moriya}.
\section{Self-Consistent Renormalization}
To cure the infrared divergences we have to determine the critical parameter $\delta_2$
self-consistently including the effect of the 4th order terms Eq.(\ref{4thorder}).
The self-consistent  equation in real space for the {\it renormalized} $\delta_2^{R}(T)$ is
\cite{moriya,nagaosa}
\be
\label{scr1}
\delta_2^{R}(T)=\delta_2+\frac{\lambda_2}{4} D_{22}(0;\delta_1, \delta_2^{R}(T))
+\frac{\lambda_3}{4} D_{11}(0;\delta_1, \delta_2^{R}(T)),
\ee
where the dependences on the critical parameter are indicated explicitly.
We can neglect the renormalization of $\delta_1$ since it does not control the critical behaviour.
Let $\delta_1^*,\delta_2^*$ be the bare value of $\delta_1, \delta_2$
 for which $\delta_2^{R}(T=0)=0$.
In other words, $\delta_1^*,\delta_2^*$ defines the zero temperature QCP.
Then, the temperature dependence of the $\delta_2^{R}(T)$ for  $\delta_1^*,\delta_2^*$ can
be determined.
By subtracting the Eq.(\ref{scr1}) at $T \neq 0$ and $T=0$ for $\delta_1^*,\delta_2^*$, we get
\be
\label{scr2}
\delta_2^{R}(T)=\frac{\lambda_2}{4}\Big( D_{22}(0;\delta_1^*, \delta_2^{R}(T))-
D_{22}(0;\delta_1^*, \delta_2^{R}=0) \Big)
+\frac{\lambda_3}{4}\Big( D_{11}(0;\delta_1^*, \delta_2^{R}(T))-
D_{11}(0;\delta_1^*, \delta_2^{R}=0) \Big).
\ee
Using the relation,
\be
D_{ij}(0)=\int \frac{d^2 \vq}{(2\pi)^2} \int_0^{\infty} \frac{d \omega}{\pi}
\coth \frac{\omega}{2T} {\rm Im} D_{ij}^R(\omega,\vq).
\ee
we can reexpress  Eq.(\ref{scr2}) as
\ba
\label{scr3}
\delta_2^{R}(T)&=&\frac{\lambda_2}{4}
\int \frac{d^2 \vq}{(2\pi)^2} \int_0^{\infty} \frac{d \omega}{2\pi}
\Big[ (\coth \frac{\omega}{2T}-1)  {\rm Im} D^R_{22,\delta_2^R(T)}(\omega,\vq) \nonumber \\
&+&\Big( {\rm Im} D^R_{22,\delta_2^R(T)}(\omega,\vq)-{\rm Im} D^R_{22,\delta_2^R(T)=0}(\omega,\vq)
\Big)\Big] \nonumber \\
&+&\frac{\lambda_3}{4} \Big[ D_{22} \to D_{11} \Big] .
\ea
The first term of Eq.(\ref{scr3}) determines the temperature dependence.
The last term of Eq.(\ref{scr3}) ($D_{11}$ part ) gives the subleading contribution proportional
to $T^3$, while the first term and the second term give
$ T \ln (\delta_2^R(T) + T^2)  + {\rm const.} \times\delta_2^R(T)$.
At low temperature, the dominant contribution to the integral comes from the low frequency
region, where the Landau damping gives the largest contribution to
${\rm Im} D^R(\omega,\vq)$. Carrying out integrals up to logarithmic accuracy, we get
\be
\delta_2^R(T) \sim T \ln \frac{1}{T}.
\ee
Now we compute some physical properties based on the self-consistent renormalization scheme.

{\bf Specific Heat}-
The singular behaviour of  specific heat near  QCP can be obtained by computing
free energy with  renormalized $\delta_2^R(T)$.
Integrating out order parameter $N$ with the effective action Eq. (\ref{quadratic}), we get the
singular piece of the free energy in (self-consistent) ladder approximation.
\be
\Delta F=\int \frac{d^2 \vq}{(2\pi)^2} \int_0^{\infty} \frac{d \omega}{\pi}
\coth \frac{\omega}{2T}
{\rm Im} \ln \Big [ \pi_1^R(\omega,\vq) \pi_2^R(\omega,\vq)-(\pi_3^R(\omega,\vq))^2 \Big].
\ee
The most dominant contribution to the free energy at low temperature
comes from the low-$\omega$ region. In that
region, we can neglect the $\omega^2$ terms of  polarization functions $\pi^R_i$
 in the free energy compared to Landau damping term.
Then, the most singular contribution comes from $\pi_2^R(\omega,\vq)$.
\be
\Delta F_{{\rm sing}} \sim \int \frac{d^2 \vq}{(2\pi)^2} \int_0^{\infty} \frac{d \omega}{2\pi}
\coth \frac{\omega}{2T}{\rm tan}^{-1}\left(\frac{\omega /\Gamma_2 |\vq|}{
\delta_2^R(T)+ a_2 |\vq|^2}\right),
\ee
Using the expansion $\tan^{-1} y \sim y-y^3/3$, and taking  derivative of
$\Delta F_{{\rm sing}}$ with respect to
temperature we get the singular part of the specific heat.
\be
\label{specificheat}
C_v \sim  \frac{T}{\sqrt{\delta_2^R(T)}}- \frac{T^3}{T (\delta_2^R(T))^3}.
\ee
The above result is {\it not} compatible with experimental data in normal state which shows
almost T-linear behaviour. We have to keep in mind that the result Eq.(\ref{specificheat}) is
essentially due to the intra-band contribution to the polarization function $\pi_2^R$.
In three dimension, the above result would be
\be
C_v \sim  T \ln \delta_2^R(T)- \frac{T^3}{(\delta_2^R(T))^3} \ln \frac{T}{\sqrt{\delta_2^R(T)}}.
\ee

{\bf Electron Self-Energy}-The coupling between  order parameter and the
anti-bonding electron (conduction electron) can be obtained by substituting Eq.(\ref{transform})
into Eq.(\ref{partition}).
\be
-V a^\dag_{k^\prime} \,\Big[ N_1(k^\prime-k) f(k,k^\prime)+i N_2(k^\prime-k) g(k,k^\prime) \Big]\,
a_k,
\ee
where the coupling matrix elements are(notations are defined in Appendix A.)
\be
f/g=\frac{2 t}{(\xi_A)_k (\xi_A)_{k^\prime}}\,\Big[ E_A(\vk^\prime)(s_x^2(\vk)-s_y^2(\vk))
\pm
E_A(\vk)(s_x^2(\vk^\prime)-s_y^2(\vk^\prime))\Big].
\ee
Note that when the momentum transfer $\vk^\prime -\vk$ is small
the matrix element $g(k,k^\prime)$ is
proportional to $(\vk-\vk^\prime)\cdot(\vk+\vk^\prime)$, therefore, the coupling with
the critical order parameter $N_2$ is suppressed
for the forward scatterings.
Considering the scattering with $N_2$, the imaginary part of the electron
self-energy can be written as
\ba
\label{self1}
{\rm Im} \Sigma^R(\epsilon,\vp)&=&V^2 \int \frac{d^2 \vq}{(2\pi)^2} \int_0^\infty \frac{d
\omega}{\pi} [(\vp+\frac{\vq}{2})\cdot \vq]^2 {\rm Im} D^R_{22}(\omega,\vq) \nonumber \\
&\times&\Big[ \delta(\epsilon+\omega-\xi^A_{\vp+\vq}) (n_B(\omega)+n_F(\epsilon+\omega))
+ \delta(\epsilon-\omega-\xi^A_{\vp+\vq}) (1+n_B(\omega)-n_F(\epsilon-\omega))\Big],
\ea
where the unimportant factors of the couplings are omitted.
At zero temperature ($\delta_2^R=0$), the self-energy (\ref{self1}) becomes
(assuming $\epsilon > 0$),
\be
\label{self2}
{\rm Im} \Sigma^R(\epsilon,\vp)=V^2 \int \frac{d^2 \vq}{(2\pi)^2} \int_0^\epsilon \frac{d
\omega}{\pi} [(\vp+\frac{\vq}{2})\cdot \vq]^2 {\rm Im} D^R_{22}(\omega,\vq;\delta^R_2=0)
\delta(\epsilon-\omega-\xi^A_{\vp+\vq}).
\ee
Let us first consider the zero temperature case.
$ {\rm Im} D^R_{22}(\omega,\vq;\delta^R_2=0)$ behave differently depending on the relative magnitude
of $\omega$ and $\Gamma_2 |\vq|$ (see Eq.(\ref{d22}).).
If $\omega > \Gamma_2 |\vq|$,  then
the typical values of momentum and frequency are given by
$\omega\sim q_\parallel \sim q_\perp$.
$q_{\parallel / \perp}$ is the component of $\vq$ parallel/perpendicular to $\vp$.
Therefore, the $|\vq|^2$ term in the
vertex can be neglected compared to $\vp\cdot  \vq$. Carrying out the integrals, we find the self-energy
 is proportional to $\epsilon^2$.
If $\omega < \Gamma_2 |\vq|$, $\omega \sim q_\parallel $, while $q_\perp \sim \omega^{1/3}$. Therefore,
$q_\perp \gg q_\parallel$, the $\vp\cdot  \vq$ term in the
vertex can be neglected compared to $|\vq|^2$.
Carrying out integrals, with the momentum cut-off of the order of the external momenta $\vp$, we find the
self-energy
is proportional to $-\epsilon^2 \ln \epsilon^{1/3}$. The contribution from the region
$|\omega| < \Gamma |\vq|$ is larger than that of from $|\omega| > \Gamma |\vq|$ by a logarithmic factor.
Thus, the self-energy at zero temperature is
\be
{\rm Im} \Sigma^R(\epsilon,\vp,T=0) \sim -\epsilon^2 \ln |\epsilon|^{1/3}.
\ee
Because the vertex $[(\vp+\frac{\vq}{2})\cdot \vq]^2$ suppresse the low momentum processes very strongly, the self-energy does not
show anomalous behaviour except for the logarithmic factor.
At finite temperature,  the bose factor $n_B(\omega)$ dominates the integral. Therefore, the
frequency integral is effectively cut off by the temperature. Carrying out the integrals,
we find the result is proportional to $-T^2 \ln T^{1/3}$.
Combining the results for the zero temperature and the finite temperature cases, we can write
\be
\label{selfenergy}
{\rm Im} \Sigma^R(\epsilon,\vp,T) \sim -(\max(\epsilon,T))^2 \ln [\max(\epsilon,T)]^{1/3}.
\ee
The self-energy Eq.(\ref{selfenergy}) is like that of Fermi-liquid apart from the minor logarithmic factor.
Eq.(\ref{selfenergy}) is also essentially determined by {\em intra-band} contribution to
the polarization function like specific heat.

The results obtained within self-consistent renormalizaton Eq.(\ref{specificheat}, \ref{selfenergy})
are not compatible with those of MFL theory. Clearly, we have to incorporate the higher order corrections
beyond ladder approximation.
However, the results  Eq.(\ref{specificheat}, \ref{selfenergy})
might be applicable to overdoped region,
namely the deep  in the disordered phase, where the critical fluctuation is less important, so that
the higher order correction is not important. In this region, the temperature dependence of
$\delta_2^R(T)$ can be ignored, and the results Eq.(\ref{specificheat}, \ref{selfenergy}) are consistent with
the experimental data in overdoped region.
\section{Summary and Concluding Remarks}
In summary, we have studied a model of HTSC proposed by C. M. Varma
which predicts QCP associated with the
{\it circulating current} order parameter.
The results derived with the ladder approximation
indicates that the system is like Fermi liquid  near the quantum critical point and
in the disordered regime.  In particular, the imaginary part of electron
self-energy is proportional to
$[\max(\epsilon,T)]^2$ up to minor logarithmic correction. Thus, the ladder approximation is
not sufficient in describing the properties
anomalous normal state, for which MFL theory gives adequate phenomenological description.
For the proper explanation of the
anomalous normal state properties the
consideration of higher order corrections beyond ladder approximation seems to be crucial.

To get some clues on the role of higer order correction, let us estimate a higher order digram
 which corresponds to the vertex correction of the susceptibility $D_{22}(i\omega,\vq)$  beyond
ladder approximation.
Let us consider the renormalization of the susceptibility
by the 4th order terms of (\ref{action1}).
The full momentum and frequency dependences of $\Pi_C(q_i), \Pi_D(q_i),\Pi_E(q_i)$
should be retained
in evaluating the higher order correction. As has been discussed in Sec.IV, the coupling between
the anti-bonding conduction electrons and the order parameter is strongly suppressed for small
momentum transfer.
As can be shown directly by substituting (\ref{transform}) into (\ref{partition}) and (\ref{theM}),
the coupling $ a^\dag_{\vp+\vq} b_\vp N_2(\vq)+{\rm H.C.} $ is {\it not} suppressed at small
$\vq$.  Thus, we have to consider a diagram which renormalize $D_{22}$ with only
($ a^\dag_{\vp+\vq} b_\vp N_2(\vq)+{\rm H.C.}$)-type vertex. The first of such diagram is
(See Fig.1),
\begin{figure}[b]
\inseps{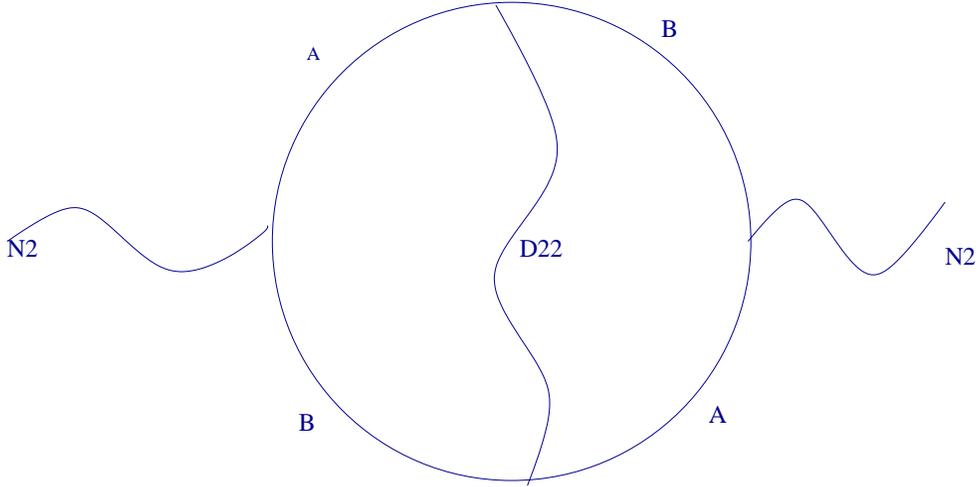}{0.66}
\caption{\label{fig1}
A higher order diagram which renormalize the susceptibility
$\langle N_2 N_2 \rangle$. The roman letter A, B indicates the Green function
$G_{A/B}$ (see Eq.(\ref{relationGreen}).), and the wiggly line is the propagator
$D_{22}(k)$.}
\end{figure}
\noindent
With the notations $q=(i\omega,\vq), p=(i\epsilon,\vp),
k=(i\Omega,\vk)$,  the analytic expression of Fig. 1 is \be ({\rm
Fig.1}) \sim V^4 \sum_{p,k}\,G_B(p) G_A(p+k) G_B(p+k+q) G_A(p+q)
D_{22}(k), \ee where the non-singular parts of vertices are
omitted. The most singular  contribution coming from the low
energy region can be estimated as follows: First,  put $k=0$ in
$G_{A/B}$ and do $p$ integral. Second, carry out the remaining
integral in $k$. If any infrared divergence occurs, cut the
divergence by $\max(q,\delta_2)$. The result of $p$ integral is
\be I_p \sim
\frac{1}{\omega^2}\,\ln\Big[1+\frac{\omega^2}{(E_B(k_F)-\mu)^2}\Big].
\ee Major contribution in momentum integral comes from  $|\vp| >
p_F$, where $\vq$ can be neglected. Usually $\omega^2 \ll
(E_B(k_F)-\mu)^2$, then $I_p$ is {\it non-singular}. The diagram
Fig.1 gives non-singular contribution, which just renormalizes
some coefficients of the susceptibility. Therefore, the
straightforward perturbative computation at two loop order does
{\it not} give indications of singular behaviour. However, if we
{\it assume}, following Varma's heuristic derivation
\cite{varma1}, that in the {\it strong} coupling limit (large
$V$) $E_B(k_F)-\mu$ can be replaced by an excitonic energy scale
which corresponds $\delta_2 $ in our notation, then $I_p$ {\it
becomes} highly singular as we approach the quantum critical point
$(\delta_2 \rightarrow 0)$. The justification of the above
replacement and the further investigation of consequences of the
singularity will be further studied. The approach using the
(non-perturbative) dynamical mean-field theory\cite{dmft} is
currently under investigation.

\bigskip

\centerline{{\bf ACKNOWLEDGEMENTS}} We are grateful to Chandra
Varma for useful discussions and comments on the manuscript. We
are also thankful to Yunkyu Bang for discussions on the three-band
Hubbard models. This work was supported by the Korea Science and
Engineering Foundation (KOSEF) through the grant No.\
1999-2-11400-005-5, and by the Ministry of Education through
Brain Korea 21 SNU-SKKU Program.

\appendix
\section{Diagonalization of Kinetic Term}
The fermion Green functions can be obtained by inverting the kernel matrix $K_0^{-1}$ Eq.(\ref{kernel}).
The diagnonalization of the free Hamiltonian $H_0$ gives three bands:
the anti-bonding $E_A(\vk)$, the bonding $E_B(\vk)$,
and the non-bonding $E_C(\vk)$.
In the limiting case $t^\prime=x t_{pp}=0$, the closed expression of each band  is
\be
E_{A,B}(\vk)=\frac{\epsilon_d}{2}\pm \frac{E(\vk)}{2},\;
E_C(\vk)=0,
\ee
where $E(\vk)=\Big[ (\epsilon_d)^2+( 4 t)^2  s^2(\vk) \Big]^{1/2}$ with
 $s^2(\vk)=s_x^2(\vk)+s_y^2(\vk)$. Finite $t^\prime$ gives dispersion to the non-bonding band which
 we ignore. However, $t^\prime$ plays a crucial role in the description of the current pattern of
 the ordered phase. Since only the disorded phase is considered in this paper,  we will take
 $t^\prime=0$.
Let $U$ be the unitary matrix which diagonalizes $K_0^{-1}$.
\be
U^\dag K_0^{-1} U = \pmatrix{ E_A(\vk)-\mu-i\epsilon & 0 & 0 \cr
                               0 & E_B(\vk)-\mu-i\epsilon & 0  \cr
                   0  & 0   & E_C(\vk)-\mu-i\epsilon \cr}.
\ee
The explicit form of the unitary matrix is given by
$U=( | {\bf r}_1 \rangle, | {\bf r}_2 \rangle, | {\bf r}_3 \rangle )$,
where $|{\bf r}_j \rangle$ is the j-th eigen-column vector of $K_0^{-1}$.
The matrix of  fermion Green's function is
\be
\hat{G}=U \pmatrix{ \frac{1}{i\epsilon+\mu-E_A(\vk)} & 0 & 0 \cr
                     0 & \frac{1}{i\epsilon+\mu-E_B(\vk)} & 0 \cr
             0  & 0 & \frac{1}{i\epsilon+\mu-E_C(\vk)} \cr} U^\dag,
\ee
where the row and the column of $\hat{G}$ is indexed by $d,p_x,p_y$.
The explicit form of the unitary matrix $U$
\be
U=\pmatrix{ \frac{ E_A(\vk)}{\xi_A(\vk)} & \frac{ E_B(\vk)}{\xi_B(\vk)} & 0 \cr
 \frac{2 t s_x(\vk)}{\xi_A(\vk)} & \frac{2 t s_x(\vk)}{\xi_B(\vk)} & \frac{s_y(\vk)}{\xi_C(\vk)} \cr
 \frac{2 t s_y(\vk)}{\xi_A(\vk)} & \frac{2 t s_y(\vk)}{\xi_B(\vk)} & \frac{-s_x(\vk)}{\xi_C(\vk)} \cr},
 \ee
where $\xi_{A/B}(\vk)=\Big[E_{A/B}^2(\vk)+(2t)^2 s^2(\vk) \Big]^{1/2}$.
When $\epsilon_d=0$, the matrix elements of $U$ simplify considerably.
\be
U=\pmatrix{ \frac{1}{\sqrt{2}} & -\frac{1}{\sqrt{2}} & 0 \cr
             \frac{s_x}{\sqrt{2} s}  & \frac{s_x}{\sqrt{2} s} & \frac{s_y}{s} \cr
         \frac{s_y}{\sqrt{2} s}  & \frac{s_y}{\sqrt{2} s} & -\frac{s_x}{s} \cr}.
\ee
The relation between the diagonalized fermion operators
$a, b, c$ and the original $(d,p_x,p_y)$ operators
are
\be
\label{transform}
[a,b,c]^{{\rm tr}}=U^\dag [d,p_x,p_y]^{{\rm tr}},
\ee
where ${\rm tr}$ denotes the tranpose of a matrix.
In case $\epsilon_d=0$, the direct relations between  fermion Green functions are
\ba
\label{relationGreen}
G_{dd}&=&\frac{1}{2}(G_A+G_B), G_{d p_x}=\frac{ s_x}{2 s}   (G_A-G_B),
G_{d p_y}=\frac{ s_y}{2 s}  (G_A-G_B), \nonumber \\
G_{p_x p_x}&=& \frac{s_x^2}{2 s^2}(G_A+G_B) + \frac{s_y^2}{s^2} G_c,
G_{p_y p_y}= \frac{s_y^2}{2 s^2}(G_A+G_B) + \frac{s_x^2}{s^2} G_c,
G_{p_x p_y}=\frac{s_x s_y}{2 s^2} (G_A+G_B- 2 G_c),
\ea
where $G_{A/B/C}=\frac{1}{i\epsilon+\mu-E_{A/B/C}(\vk)}$.
\section{Polarization Functions}
The explicit expressions of the polarization functions $\Pi_a(q)$ are
\ba
\Pi_A(q)&=&\frac{2 V^2 T}{N_0} \sum_\epsilon  \sum_\vp\,
G_{dd}(p+q)\,\Big( s_x^2(\vp)\,G_{p_x p_x}(p)+s_y^2(\vp)\,G_{p_y p_y}(p)-
2 s_x(\vp) s_y(\vp)\,G_{p_x p_y}(p) \Big) \nonumber \\
\Pi_B(q)&=&\frac{ V^2 T}{N_0} \sum_\epsilon  \sum_\vp\,
\Big[ G_{d p_x}(p) s_x(\vp)-G_{d p_y}(p) s_y(\vp)\Big]
\Big[  G_{d p_x}(p+q) s_x(\vp+\vq)-G_{d p_y}(p+q) s_y(\vp+\vq)\Big] \nonumber \\
\Pi_C(q_i)&=&3 V^4\,
\sum_p\, G_{dd}(p_1) \,G_{dd}(p_3)\,
\Big(
s_x^2(\vp_2) G_{p_x p_x}(p_2)+s_y^2(\vp_2) G_{p_y p_y}(p_2)-
2 s_x(\vp_2) s_y(\vp_2)\,G_{p_x p_y}(p_2) \Big), \nonumber \\
&\times&\Big(
s_x^2(\vp_4) G_{p_x p_x}(p_4)+s_y^2(\vp_4) G_{p_y p_y}(p_4)-
2 s_x(\vp_4) s_y(\vp_4)\,G_{p_x p_y}(p_4) \Big), \nonumber \\
\Pi_D(q_i)&=&2 V^4\,\sum_p\,G_{dd}(p_1)\,
\Big(
s_x^2(\vp_2) G_{p_x p_x}(p_2)+s_y^2(\vp_2) G_{p_y p_y}(p_2)-
2 s_x(\vp_2) s_y(\vp_2)\,G_{p_x p_y}(p_2) \Big) \nonumber \\
&\times&\Big[ G_{d p_x}(p_3) s_x(\vp_3)-G_{d p_y}(p_3) s_y(\vp_3)\Big]
\Big[  G_{d p_x}(p_4) s_x(\vp_4)-G_{d p_y}(p_4) s_y(\vp_4)\Big], \nonumber \\
\Pi_E(q_i)&=& \frac{V^4}{2}\,
\prod_{i=1,4}\,
\Big[G_{d p_x}(p_i) s_x(\vp_i)-G_{d p_y}(p_i) s_y(\vp_i) \Big],
\ea
where the signs of  momentum and  energy along the loop of the 4th order diagrams
 are to be implicitly understood.   $N_0$ is the number of lattice sites.
Let us assume $\epsilon_d=0$. Using the relation  (\ref{relationGreen}) we can compute
$\Pi_A(q)$ and $\Pi_B(q)$.
\ba
\Pi_A(q)&=& V^2 \int \frac{d^2 \vp}{(2\pi)^2} T \sum_\epsilon  \Big[
\frac{(s_x^2(\vp)-s_y^2(\vp))^2}{2 s^2(\vp)} G_A(p+q) G_A(p) \nonumber \\
&+& \frac{(s_x^2(\vp)-s_y^2(\vp))^2}{2 s^2(\vp)} \big(G_A(p+q) G_B(p)+G_A(p) G_B(p+q) \big)
\nonumber \\
&+& \frac{(s_x^2(\vp)-s_y^2(\vp))^2}{2 s^2(\vp)} G_B(p+q) G_B(p) +
\frac{ 4 s_x^2 (\vp) s_y^2 (\vp)}{s^2(\vp)} G_B(p+q) G_C(p) \nonumber \\
&+& \frac{ 4 s_x^2 (\vp) s_y^2 (\vp)}{s^2(\vp)} G_A(p+q) G_C(p) \Big].
\ea
\ba
\Pi_B(q)&=& V^2 \int \frac{d^2 \vp}{(2\pi)^2} T \sum_\epsilon
\frac{(s_x^2(\vp)-s_y^2(\vp))^2}{2 s^2(\vp)} \frac{(s_x^2(\vp+\vq)-s_y^2(\vp+\vq))^2}{2 s^2(\vp+\vq)}
\nonumber \\
&\times&\Big[ G_A(p) G_A(p+q)+G_B(p) G_B(p+q)- G_A(p) G_B(p+q)- G_B(p) G_A(p+q)\Big].
\ea
The angular factors do not introduce any singular features in the integral, and they will be ignored.
The calculation of $G_A G_A$ correlation function gives the well-known result (at zero temperature),
\be
\label{loop0}
\int \frac{d^2 \vp}{(2\pi)^2} T \sum_\epsilon G_A(p) G_A(p+q) \sim
-N(E_F) \Big[1-\frac{ |\omega|}{ v_F |\vq|}- |\vq|^2 \Big],
\ee
where $N(E_F)$ is the unrenormalized density of states at the Fermi level, and
the purely numerical constants are suppressed.
The result (\ref{loop0}) is valid for $|\vq|< p_F, |\omega| < v_F |\vq|$.
In the regime, $ |\omega| \gg v_F |\vq|$, the correlation function (\ref{loop0}) is proprotional to
$|\vq|^2/\omega^2$, therefore, it is small.
The summation over frequency of  $G_A(p+q) G_B(p)$ correlation function gives
\be
\label{loop1}
\int \frac{d^2 \vp}{(2\pi)^2} T \sum_\epsilon G_A(p+q) G_B(p)=\int \frac{d^2 \vp}{(2\pi)^2}
\frac{n_F(E_B(\vp)-\mu)-n_F(E_A(\vp+\vq)-\mu)}{i\omega+E_B(\vp)-E_A(\vp+\vq)}.
\ee
Similary,
\be
\label{loop2}
\int \frac{d^2 \vp}{(2\pi)^2} T \sum_\epsilon G_B(p+q) G_A(p)=\int \frac{d^2 \vp}{(2\pi)^2}
\frac{n_F(E_A(\vp)-\mu)-n_F(E_B(\vp+\vq)-\mu)}{+i\omega+E_A(\vp)-E_B(\vp+\vq)}.
\ee
At low temperature $n_F(E_B(\vp)-\mu) \approx 1$, therefore, the anti-bonding electrons should
lie outside the Fermi surface. Thus, for the momentum $|\vq|< p_F$, (\ref{loop1}) and (\ref{loop2}) do
not depend on $\vq$. Carrying out the integrals, we get for $|\omega| < 2 t$,
\be
\int \frac{d^2 \vp}{(2\pi)^2} T \sum_\epsilon \Big(G_B(p) G_A(p+q) + G_A(p) G_B(p+q) \Big)
\sim -\frac{\Lambda_0-p_F}{4\pi t} \Big(1- \frac{\omega^2}{(2t)^2} \Big).
\ee
$\Lambda_0$ is the momentum cut-off, or the size of Brillouin zone.
In the opposite limit, $|\omega| \gg 2 t$,
\be
\int \frac{d^2 \vp}{(2\pi)^2} T \sum_\epsilon \Big(G_B(p) G_A(p+q) + G_A(p) G_B(p+q) \Big)
\sim -\frac{\Lambda_0-p_F}{2\pi } \frac{2t}{|\omega|^2}.
\ee
$G_A G_C$ can be computed analogously.
\be
\int \frac{d^2 \vp}{(2\pi)^2} T \sum_\epsilon G_c(p) G_A(p+q) \sim
\frac{N(E_F)}{2\pi}\,\ln \frac{\mu-i \omega}{\xi_0+\mu},
\ee
where $\xi_0$ is the energy cut-off.
In the limit, $|\omega| < \mu$,
\be
\int \frac{d^2 \vp}{(2\pi)^2} T \sum_\epsilon G_c(p) G_A(p+q) \sim
-\frac{N(E_F)}{2\pi} \Big({\rm const.}+\frac{i \omega}{\mu} \Big).
\ee
The correlation functions $G_B G_B$ and $G_B G_C$ give negligible contributions at low temperature
because the bonding band and the non-bonding band are fully occupied.
Combining all of the above results, we can write
\ba
\label{polarization}
\Pi_A(q)&=& b_1 N(E_F) V^2 \Big[-c_1+ |\vq|^2 + \frac{|\omega|}{v_F |\vq|}+  \frac{\omega^2}{(2t)^2}
+ \ln \frac{\mu-i \omega}{\xi_0+\mu}\Big], \nonumber \\
\Pi_B(q)&=& b_2 N(E_F) V^2 \Big[c_2+|\vq|^2 + \frac{|\omega|}{v_F |\vq|}-\frac{\omega^2}{(2t)^2}\Big],
\ea
where $b_1,b_2, c_1$ are the {\it positive} numerical constants.  $c_2$ is also a numerical constants
which can positive or negative
depending on doping and bandwidth. In case of our interest, $c_2$ is positive.


\begin{references}
\bibitem{comment1} Hereafter, unless indicated explicitly, the lattice unit $a=1$ will be used
throughout this paper.
\bibitem{comment2} $x_c \sim 0.2$ is an estimate obtained from the mean-field theory.
\end{references}
\end{document}